\documentclass[preprint,preprintnumbers,amsmath,amssymb]{revtex4}

\usepackage{graphicx}
\usepackage{dcolumn}
\usepackage{bm}

\begin{document}

\preprint{MIT-CTP-3318}

\title{An Index Theorem for Domain Walls in Supersymmetric Gauge Theories}

\author{Keith S. M. Lee}
 \email{ksml@mit.edu}
\affiliation{%
 Center for Theoretical Physics,\\
Massachusetts Institute of Technology, \\ Cambridge, MA 02139, U.S.A.\\
}%

\date{\today}

\begin{abstract}
The supersymmetric abelian Higgs model with $N$ scalar fields admits 
multiple domain wall solutions. We perform a Callias-type index 
calculation to determine the number of zero modes of this 
soliton. We confirm that the most general domain wall has 
$2(N-1)$ zero modes, which can be interpreted as the 
positions and phases of $(N-1)$ constituent domain walls. This 
implies the existence of moduli for a $D$-string interpolating 
between $N$ $D5$-branes in $IIB$ string theory.
\end{abstract}

\maketitle

\section{\label{sec:intro} Introduction}

There exists an intimate connection between supersymmetry and solitons
\cite{WO78}. 
In many supersymmetric theories, the second-order equations of motion collapse 
to a first-order Bogomol'nyi equation, with a moduli space of solutions. 
The physical interpretation of this moduli space is that the competing 
forces between solitons cancel.

However, in the case of domain walls, supersymmetry does not 
guarantee the absence of forces \cite{PT02}.
In fact, there is something of a conflict between having domain walls and 
maximizing supersymmetry. While the former requires isolated vacua, 
the latter typically implies an extended moduli space of vacua. 
Recently there has been much interest in models in which the forces do
cancel \cite{ms98}--\cite{PV01}. 
In this regard, it is interesting to study the unique class of maximally
supersymmetric theories admitting multiple domain wall solutions.
This is ${\cal N}=2$ supersymmetric QED, in which the introduction of a 
Fayet-Illiopoulos (FI) parameter lifts the Coulomb branch, while 
non-zero masses lift the Higgs branch. 

Domain walls in this theory have been discussed in \cite{DT02}, where it is 
conjectured that the most general domain wall solution admits $2(N-1)$ 
collective coordinates with the interpretation of the positions and phases 
of $(N-1)$ constituent domain walls. 
In the strong coupling limit, $e^2 \rightarrow \infty$, the model reduces
to a massive non-linear sigma model on the 
Higgs branch of the theory \cite{ew93}, and has a surprisingly rich 
spectrum of solitons \cite{ea92}--\cite{NNS02}, \cite{DT02}. 
In this case, the theory contains only scalar fields,
and so one can use Morse theory to determine the number of moduli of the
domain walls \cite{ew82}. This  calculation was performed in \cite{GTT01b}, 
where it was found that there are indeed no forces and a moduli space of
solutions exists. 

However, for finite gauge coupling constant, the presence of the gauge field
means that Morse theory techniques are not applicable.
The purpose of this paper is to examine whether domain wall zero modes remain 
in the general case. 
We adapt index theorems of Weinberg (who followed a procedure developed by
Callias \cite{cjc78}), and confirm that
the system retains all $2(N-1)$ zero modes.

This theory finds application in the $D1$-$D5$ system of string theory 
in the presence of a background NS-NS $B$-field. The domain wall 
describes a $D$-string interpolating between separated $D5$-branes 
\cite{BT99,LT00}. 
The index theorem presented here confirms that the $D$-string has 
moduli. 

We shall calculate the index by the introduction of a regulator 
scale, $M$. As can be seen from (\ref{eq:indMres}), the final answer 
is dependent on $M$. This is in contrast to the index for instantons 
and vortices, but also occurs for monopoles \cite{ejw79}. In 
the latter case, the $M$-dependence is related to both the quantum mass 
renormalization of monopoles \cite{rkk84} in four dimensions, 
and the non-cancellation of determinants in three-dimensional 
instanton calculations \cite{DKMTV97}. In the present case, our 
calculation also contains similar information regarding the 
mass renormalization of kinks in $(1+1)$-dimensional field theories, 
and the one-loop effects in the background of instantons 
in supersymmetric quantum mechanics.

The rest of the paper is organized as follows.
In Sec.~II, we review the gauge theory under consideration and 
its domain wall solutions, and in Sec.~III, we calculate the 
dimension of the moduli space.

\section{\label{sec:SQED} Gauge theory domain walls}

The theory under consideration is $d=3+1$, ${\cal N}=2$ supersymmetric 
$U(1)$ gauge theory coupled to $N$ hypermultiplets. The bosonic 
part of the Lagrangian is given by:
\begin{eqnarray}
{\cal L}&=&\frac{1}{4e^2}F^2+\frac{1}{2e^2}|\partial\phi|^2 
+\sum_{i=1}^N\left(|{\cal D}q_i|^2+|{\cal D}\tilde{q}_i|^2 
\right) -\sum_{i=1}^N|\phi-m_i|^2(|q_i|^2+|\tilde{q}_i|^2) \nonumber\\ 
&& - \frac{e^2}{2}(\sum_{i=1}^N|q_i|^2-|\tilde{q}_i|^2
-\zeta)^2-\frac{e^2}{2}|\sum_{i=1}^N\tilde{q}_iq_i|^2.
\label{lag}
\end{eqnarray}
The scalar field $q_i$ ($\tilde{q}_i$) has charge $+1$ ($-1$) under the gauge
group and complex mass $m_i$, whereas $\phi$ is a neutral complex scalar field.
The FI parameter $\zeta$ must be non-zero to lift the Coulomb branch and
can be taken to be positive, without loss of generality. 

As in \cite{DT02}, we consider the case of non-zero,  distinct masses: 
$m_i\neq m_j$ for $i\neq j$. We take these to be real and consequently 
can choose the ordering 
$
m_{i+1}<m_i \; \hbox{for all} \; i.
   \label{eq:morder}
$
There are then $N$ isolated vacua, given by:
\begin{equation}
  \hbox{Vacuum} \; i: \;\; \phi=m_i, \;  |q_j|^2=\zeta\delta_{ij}, \;\;
  |\tilde{q}_j|^2=0.  
     \label{eq:vacua}
\end{equation}
Furthermore, certain fields do not appear in the domain wall solutions
and so are set to zero now:
$
{\rm Im}\,(\phi)=\tilde{q}_i=F=0.
\label{set}
$
Thus, the field $\phi$ is real.

We choose the domain walls to be oriented in the $x^2-x^3$ plane, in which case
$\partial_2 = \partial_3 \equiv 0$. 
By completing the square in the Hamiltonian, one finds that
the Bogomol'nyi equations that minimize the potential energy are given by:
\begin{eqnarray}
\partial\phi&=&e^2(\sum_{i=1}^N|q_i|^2-\zeta) \label{bog1} \\ 
{\cal D}q_i&=&(\phi-m_i)q_i. \label{bog2}
\end{eqnarray}
Here $\partial \equiv \partial_1$ and ${\cal D} = \partial - i A$, where 
$A \equiv A_1$ is the gauge potential.
 
In what follows, we shall concentrate upon domain walls interpolating between 
the first and $N$th vacua. It is conjectured that these kinks will decompose 
into many kinks, each interpolating between vacuum $i$ and vacuum $i+1$. 
We wish to investigate this.

\section{\label{sec:index} The dimension of the domain wall moduli space}

From (\ref{bog1}) and (\ref{bog2}), the linearized Bogomol'nyi equations are:
\begin{eqnarray}
\partial\dot{\phi}&=&e^2\sum_{i=1}^N(\dot{q}_i{q}^\dagger_i+q_i
\dot{q}^\dagger_i) \label{lbog1}\\
{\cal D}\dot{q}_i-i\dot{A}q_i&=&(\phi-m_i)\dot{q}_i+\dot{\phi}q_i,
\label{lbog2}
\end{eqnarray}
in which dots represent small fluctuations in the fields, i.e.\
$\dot{\phi} \equiv \delta\phi$.
These must be supplemented with a gauge-fixing condition. A convenient
choice which is compatible with supersymmetry is provided by Gauss' law in
$A_0=0$ gauge:
\begin{eqnarray}
\partial \dot{A}=ie^2\sum_{i=1}^N(q_i \dot{q}^\dagger_i
-{q}^\dagger_i \dot{q}_i).
\label{gauss}
\end{eqnarray}
Then (\ref{lbog1}) and (\ref{gauss}) can be combined
into:
\begin{equation}
\partial(\dot{\xi})=2e^2\sum_{j=1}^N\dot{q}_jq_j^\dagger,
   \label{lbogauss}
\end{equation}
where we have defined $ \xi = \phi + i A $.

Writing (\ref{lbogauss}) and (\ref{lbog2}) in matrix form, we have:
\begin{equation}
    \mathbb{D} \Psi = 0, 
\end{equation}  
where
$
   \Psi = (\dot{\xi}, \dot{q}_1, \dot{q}_2, \ldots, \dot{q}_N)^{T} 
$
and
\begin{equation}
     \mathbb{D}
   = \left(\begin{array}{ccccc}
    \partial & -2e^2 q_1^\dagger & -2e^2 q_2^\dagger  
             & \cdots  & -2e^2 q_N^\dagger \\  
    -q_1 & \partial - (\xi - m_1)  & 0   & \cdots & 0 \\
    -q_2 & 0 &  \partial - (\xi - m_2) &  & \\  
     \vdots & \vdots &  & \ddots & \\
    -q_N & 0  &   &  & \partial - (\xi - m_N)
   \end{array}\right),
   \label{eq:D}
\end{equation}
in which only entries in the first column, first row and main diagonal
are non-zero.

Our task, then, is to determine the dimension of the kernel of 
$\mathbb{D}$, which we do by means of an index theorem.
Previously, Weinberg has used index theorems to count
parameters of multivortex solutions in Ginzburg-Landau theory
\cite{ejw79a} and multimonopole solutions, first in $SU(2)$ gauge theory and 
later generalizing to an arbitrary compact simple gauge group 
\cite{ejw79,ejw80}.
The index of $\mathbb{D}$ is given by:
\begin{equation}
  {\cal I} = \lim_{M^2 \rightarrow 0} {\cal I}(M^2), 
\end{equation}
where
\begin{equation}
  {\cal	I}(M^2) 
  = \hbox{Tr} \left( \frac{M^2}{\mathbb{D}^\dagger \mathbb{D} + M^2} 
              \right)
  - \hbox{Tr} \left( \frac{M^2}{\mathbb{D} \mathbb{D}^\dagger + M^2} 
              \right).
   \label{eq:indexM}
\end{equation}
Since $\hbox{ker}(\mathbb{D}) = \hbox{ker}(\mathbb{D}^\dagger \mathbb{D})$ 
and  
$\hbox{ker}(\mathbb{D}^\dagger) = \hbox{ker}(\mathbb{D} \mathbb{D}^\dagger)$,
each zero mode of $\mathbb{D}$ contributes $1$ to the index,
while each zero mode of $\mathbb{D}^\dagger$ contributes $-1$.
If the continuum parts of the spectra extend to zero, then there is 
potentially a contribution from this source: this complication does not 
arise here because the theory has a mass gap.
Hence, if there are no zero modes of $\mathbb{D}^\dagger$, then the index 
equals the number of collective coordinates. We now show that this is
indeed the case. Consider the kernel of $ \mathbb{D}^\dagger $. 
If
\begin{equation}
    \mathbb{D}^\dagger \Psi = 0, 
\end{equation}
then (\ref{eq:D}) implies that the components of $\Psi$ satisfy:
\begin{eqnarray}
\partial \psi_0 + \sum_{i=1}^{N} q_{i}^{\dagger} \psi_i & = 0 
   \nonumber \\
2e^2 q_j \psi_0 + (\partial + (\xi^\dagger - m_j))\psi_j & = & 0, \;\;
  j = 1, 2, \ldots, N.
\end{eqnarray}
Thus,
\begin{eqnarray}
0 & = &  \int dx \left( 
      2e^2 |\partial \psi_0 + \sum_{i=1}^{N} q_{i}^{\dagger} \psi_i |^2
    + \sum_{j=1}^{N} 
      | 2e^2 q_j \psi_0 + (\partial + (\xi^\dagger - m_j))\psi_j |^2
          \right)
   \nonumber \\
  & = &  \int dx \left( 2e^2 | \partial \psi_0 |^2 
       + 2e^2 \sum_{i=1}^{N} |q_i|^2 |\psi_i|^2
       + 4e^4 \sum_{j=1}^{N} |q_j|^2 |\psi_0|^2
    \right. \nonumber \\
  & & \;\; \left.
      \quad\quad 
       + \sum_{j=1}^{N} | (\partial + (\xi^\dagger - m_j))\psi_j) |^2
   \right),
\end{eqnarray}
where we have used integration by parts and the second Bogomol'nyi
equation (\ref{bog2}) to show that the cross-terms disappear.
Hence, $\psi_i = 0, \;\; i = 0, 1, \ldots, N$, and the kernel of
$\mathbb{D}^\dagger$ is trivial. 

Note that if ${\cal I}(M^2)$ were independent of $M$, then $M$ could be chosen
to be $M \rightarrow \infty$ to simplify calculations. However, this is the 
case only if physical fields fall sufficiently rapidly at spatial infinity. 
Yang-Mills instantons and vortices are in this category, whereas
monopoles are not (see \cite[Appendix A]{ejw79} for further discussion).
Like monopoles, the kink solutions under consideration have an $M$-dependent 
index.

Following the method of \cite{ejw81}, let us define:
\begin{equation}
   \Theta = \left(\begin{array}{cc} 
             0              &  - \mathbb{D}^\dagger \\ 
             \mathbb{D}  &  0 
            \end{array}\right).
\end{equation}
Also define:
\begin{equation}
   \Gamma = \left(\begin{array}{cc}
       	     0  &  I \\
       	     I  &  0
            \end{array}\right),
   \;\;\; 
   \Gamma_5 = \left(\begin{array}{cc}
       	     I  &  0 \\
       	     0  &  -I 
            \end{array}\right).
\end{equation}
Finally, define the matrix $K(x)$ via:
\begin{equation}
  \Theta = \Gamma \partial + K(x).
   \label{eq:K}
\end{equation}
Now, ${\cal I}(M^2)$ can be re-written as:
\begin{eqnarray}
   {\cal I}(M^2) & = & \hbox{Tr} \; \Gamma_5 \frac{M^2}{-\Theta^2 + M^2}
   = \int dx \; \hbox{tr} \, \left\langle x \left| 
             \Gamma_5 \frac{M^2}{-\Theta^2 + M^2}
                           \right| x \right\rangle .
\end{eqnarray}
Let us define the nonlocal current:
\begin{equation}
   J(x,y,M)  =  \hbox{tr} \left\langle x \left| 
             \Gamma_5 \Gamma \frac{1}{\Theta + M}
       	       	       	   \right| y \right\rangle. 
\end{equation}
From (\ref{eq:K}) we have:
\begin{eqnarray}
  \delta(x-y) & = & [ \Gamma \partial_x + K(x) + M ] \left\langle x \left|  
             \frac{1}{\Theta + M} \right| y \right\rangle
               \nonumber \\
              & = & 
             \left\langle x \left| \frac{1}{\Theta + M} \right| y \right\rangle
           	[ - \overset{\leftarrow}{\partial}_y \Gamma + K(y) + M ].
\end{eqnarray}
Using this and the relations
$
  \{ \Gamma_5, K \} = \{ \Gamma_5, \Gamma \} = \{ \Gamma_5, \Theta \} 
  = 0,
    \label{eq:anticom}
$
we obtain the following:
\begin{eqnarray}
  (\partial_x + \partial_y) J(x,y,M) 
  & = & -2 \; \hbox{tr} \left\langle x \left| \Gamma_5 
             \frac{M}{\Theta + M} \right| y \right\rangle
      \nonumber \\
  & & \;\;\;\;  + \; \hbox{tr} \left( [K(x) - K(y)] \Gamma_5 
    \left\langle x \left| \frac{1}{\Theta + M} \right| y \right\rangle
                \right).
  \\
  \Rightarrow \;\;
  \partial J(x,x,M) & = & -2 \; \hbox{tr} \left\langle x \left| \Gamma_5
             \frac{M^2}{-\Theta^2 + M^2} \right| x \right\rangle.
\end{eqnarray}
Thus,
\begin{equation}
   {\cal I}(M^2) 
   = - \frac{1}{2} \left[ J(x,x,M) \right]_{x= -\infty}^{\infty}.
      \label{eq:indexM2}
\end{equation}
We shall find it more convenient to evaluate $J(x,x,M)$ from
\begin{equation}
   J(x,x,M)  =  - \hbox{tr} \left\langle x \left| 
             \Gamma_5 \Gamma \Theta \frac{1}{-\Theta^2 + M^2}
       	       	       	   \right| x \right\rangle,
      \label{eq:Jalt}
\end{equation}
which follows from the fact that 
$ \hbox{tr} (\Gamma_5\Gamma M / (-\Theta^2 + M^2)) = 0 $.

We now proceed to calculate (\ref{eq:indexM2}).
From (\ref{eq:D}), lengthy but straightforward calculations yield 
$ \mathbb{D}^\dagger \mathbb{D}$ and
$ \mathbb{D} \mathbb{D}^\dagger$. These turn out to be somewhat messy
matrices. For example, in $ \mathbb{D}^\dagger \mathbb{D}$ the entries
$(i+1, i+1)$, $(1,i+1)$ and $(i+1,j+1)$, for $1 \leq i,j \leq N$, $j \neq i$, 
are given by 
$ (-\partial^2 + 4e^4 |q_i|^2 + |\xi-m_i|^2 
         + \partial\xi - \xi^\dagger \partial)$,
$ ( 2e^2 \partial q_i^\dagger - q_i^\dagger (\partial - (\xi-m_i)) )$
and
$( 4e^4 q_i q_j^\dagger )$, respectively.

Since 
\begin{equation}
   \Gamma_5 \Gamma \Theta \frac{1}{-\Theta^2 + M^2}
 =  \left(\begin{array}{cc}
     \mathbb{D} (\mathbb{D}^\dagger \mathbb{D} + M^2)^{-1} &  0 \\
    0 & \mathbb{D}^\dagger (\mathbb{D} \mathbb{D}^\dagger + M^2)^{-1} 
            \end{array}\right),
     \label{eq:g5gt}
\end{equation}
we wish to be able to invert the matrices 
$(\mathbb{D}^\dagger \mathbb{D} + M^2)$ and
$(\mathbb{D} \mathbb{D}^\dagger + M^2)$.
Recall, though, that all that we require are the values of $J$ at $\pm \infty$ 
(corresponding to the $N$th and $1$st vacua, respectively). 
In the $i$th vacuum, the fields have values given by (\ref{eq:vacua}) and
\begin{equation}
  \partial q_j = (\xi - m_j) q_j = i A q_i \delta_{ij},
     \label{eq:vacdq}
\end{equation}
which follows from (\ref{bog2}). 
Applying these boundary conditions greatly simplifies these matrices,
each of which then has only two non-zero entries off the main diagonal.
Nevertheless, even matrices of this sparse form cannot be inverted if their 
entries do not commute. In our case, the problem arises from the presence of
the differential operator. However, we can set $A = 0$, after which
(\ref{eq:vacua}) and (\ref{eq:vacdq}) imply that 
$\partial q_j = \partial \xi = 0$ at spatial infinity.
Thus, we are able to obtain expressions for
$(\mathbb{D}^\dagger \mathbb{D} + M^2)^{-1}$ and
$(\mathbb{D} \mathbb{D}^\dagger + M^2)^{-1}$ in the first and $N$th vacua.

Finally, putting these calculations 
into (\ref{eq:indexM2})--(\ref{eq:g5gt}),
and using
\begin{eqnarray}
  \left\langle x \left| 
     \frac{m_i - m_j}{-\partial^2 + (m_i - m_j)^2 + M^2}
  \right| x \right\rangle 
   & = & \frac{m_i - m_j}{2\pi} \int \frac{dk}{k^2 + (m_i - m_j)^2 + M^2}
   \nonumber \\
   & = &  \frac{1}{2} \frac{m_i - m_j}{\sqrt{(m_i - m_j)^2 + M^2}},
\end{eqnarray}
we obtain:
\begin{eqnarray}
{\cal I}(M^2) 
   & = & \frac{1}{2} \left( 
  \sum_{i=1}^{N-1} \frac{m_i - m_N}{\sqrt{(m_i - m_N)^2 + M^2}}
+ \sum_{j=2}^{N} \frac{m_1 - m_j}{\sqrt{(m_1 - m_j)^2 + M^2}}
      \right).
   \label{eq:indMres}
\end{eqnarray}
Hence, due to the ordering $m_{i+1} < m_i$ for all $i$, 
\begin{equation}
   {\cal I} = N-1.
\end{equation}

\medskip

Therefore, we conclude that ${\cal N}=2$ SQED with $N$ hypermultiplets 
admits multi-domain wall solutions having $N-1$ complex collective coordinates.

\acknowledgments
 It is a pleasure to acknowledge the assistance of David Tong
 at all stages of this work.
 This research is supported in part by the U.S. Department of Energy 
 under cooperative research agreement \#DF-FC02-94ER40818.


\begin{thebibliography}{99} 

\bibitem{WO78}
E. Witten and D. Olive, Phys.\ Lett. {\bf 78B}, 97 (1978).

\bibitem{PT02}
R. Portugues and P. K. Townsend, Phys.\ Lett. {\bf B530}, 227 (2002).

\bibitem{ms98}
M. Shifman, Phys.\ Rev.\ D {\bf 57}, 1258 (1998);
M. Shifman and M. Voloshin,  Phys.\ Rev.\ D {\bf 57}, 2590 (1998).

\bibitem{BdR95}
D. Bazeia, M. dos Santos and R. Ribeiro,  Phys.\ Lett. {\bf A208} 84 (1995).

\bibitem{AGM02}
A. Alonso Izquierdo, M. A. Gonzalez Leon and J. Mateos Guilarte,
Phys.\ Rev.\ D  {\bf 65} 085012 (2002).

\bibitem{BHP01}
C. Bachas, J. Hoppe and B. Pioline, JHEP {\bf 0107} 041 (2001).
 
\bibitem{PV01}
L. Pogosian and T. Vachaspati, Phys.\ Rev.\ D {\bf 64}, 105023 (2001);
L. Pogosian,  Phys.\ Rev.\ D {\bf 65}, 065023 (2002). 

\bibitem{DT02}
D. Tong,  Phys.\ Rev.\ D {\bf 66}, 025013 (2002).

\bibitem{ew93}
E. Witten, Nucl.\ Phys.\ {\bf B403}, 159 (1993).

\bibitem{ea92}
E. Abraham, Phys.\ Lett. {\bf B278}, 291 (1992). 

\bibitem{AT92}
E. Abraham and P. K. Townsend, Phys.\ Lett.\ {\bf B291}, 85 (1992);
 Phys.\ Lett. {\bf B295}, 225 (1992).

\bibitem{ea93}
E. Abraham, Nucl.\ Phys.\ {\bf B399}, 197 (1993).

\bibitem{GTT01}
J. P. Gauntlett, D. Tong and P. K. Townsend,
Phys.\ Rev.\ D {\bf 63}, 085001 (2001).

\bibitem{GPTT01}
J. P. Gauntlett, R. Portugues, D. Tong and P. K. Townsend,
Phys.\ Rev.\ D {\bf 63}, 085002 (2001).

\bibitem{GTT01b}
J. P. Gauntlett, D. Tong and P. K. Townsend,  
Phys.\ Rev.\ D {\bf 64}, 025010 (2001).

\bibitem{NNS02}
M. Naganuma, M. Nitta and N. Sakai,
Grav.\ Cosmol.\ {\bf 8}, 129 (2002). 

\bibitem{ew82}
E. Witten, J.\ Diff.\ Geom.\ {\bf 17}, 661 (1982).
 
\bibitem{cjc78}
C. J. Callias, Commun.\ Math.\ Phys.\ {\bf 62}, 213 (1978).

\bibitem{BT99}
E. Bergshoeff and P. K. Townsend, JHEP {\bf 9905}, 021 (1999).

\bibitem{LT00}
N. D. Lambert and D. Tong, Nucl.\ Phys.\ {\bf B569}, 606 (2000).

\bibitem{ejw79}
E. J. Weinberg, Phys.\ Rev.\ D {\bf 20}, 936 (1979).

\bibitem{rkk84}
R. K. Kaul, Phys.\ Lett.\ {\bf B143}, 427 (1984).
 
\bibitem{DKMTV97}
N. Dorey, V. V. Khoze, M. P. Mattis, D. Tong and S. Vandoren,
Nucl.\ Phys.\ {\bf B502} 59 (1997).

\bibitem{ejw79a}
E. J. Weinberg, Phys.\ Rev.\ D {\bf 19}, 3008 (1979).

\bibitem{ejw80}
E. J. Weinberg, Nucl.\ Phys.\ B {\bf 167}, 500 (1980).

\bibitem{ejw81}
E. J. Weinberg, Phys.\ Rev.\ D {\bf 24}, 2669 (1981).

\end{thebibliography}
\end{document}